%%%%%%%%%%%%%%%%%%%%%%%%%%%%%%%%%%%%%%%%%%%%%%%%%%%%%%%%%%%%%%%%%%%%%%
%\\ 
%Title:       Steps towards Lattice Virasoro Algebras: su(1,1) 
%Authors:     M. R\"osgen, R. Varnhagen
%Comments:    15 pages, plain TeX, 4 typos corrected
%Report-no:   BONN-TH-94-23
%Journal-ref: Phys. Lett. B 350 (1995) 203 - 211
%\\
%An explicit construction is presented for the action of the su(1,1)
%subalgebra of the Virasoro algebra on path spaces for the c(2,q) 
%minimal models.
%In the case of the Lee-Yang edge singularity, we show how this action 
%already fixes the central charge of the full Virasoro algebra.
%For this case, we additionally construct a representation in terms
%of generators of the corresponding Temperley-Lieb algebra.
%\\
%%%%%%%%%%%%%%%%%%%%%%%%%%%%%%%%%%%%%%%%%%%%%%%%%%%%%%%%%%%%%%%%%%%%%%
\newcount\vers\vers=0%%%0=hepth,1=BONN-TH,2=PLB
\font\tenss=cmss10 scaled \magstep0
\font\sevenss=cmss10 at 7pt
\font\fivess=cmss10 at 5pt
\def\BNte{{\hbox{\tenss I{\hbox to 0.6em{\hss\tenss N}}}}}%<0.62
\def\BNse{{\hbox{\sevenss I{\hbox to 0.6em{\hss\sevenss N}}}}}
\def\BNfi{{\hbox{\fivess I{\hbox to 0.6em{\hss\fivess N}}}}}
\def\BN{{\mathchoice{\BNte}{\BNte}{\BNse}{\BNfi}}}
\def\BZte{{\hbox{\tenss Z{\hbox to 0.2em{\hss\tenss Z}}}}}
\def\BZse{{\hbox{\sevenss Z{\hbox to 0.2em{\hss\sevenss Z}}}}}
\def\BZfi{{\hbox{\fivess Z{\hbox to 0.2em{\hss\fivess Z}}}}}

%%%%%%%%%%%%%%%%%%%%%%%%%%%%%%%%%%%%%%%%%%%%%%%%%%%%%%%%%%%%%%%%%%%%%%%%%%
\parindent=0pt\overfullrule=0pt
\font\inbigbold=cmbx10 scaled\magstep2
\def\sn{\smallskip}\def\mn{\medskip}\def\bn{\bigskip}  
\global\newcount\kapnum\global\newcount\glgnum
\global\newcount\chapflag\global\newcount\footflag
\global\kapnum=0\global\glgnum=0\global\chapflag=0\global\footflag=0
\xdef\cph{}
\def\chaphead#1{\xdef\cph{#1}}
\def\chapter#1{{\chaphead{#1}\inbigbold\leftline{#1}}
               \global\advance\kapnum by1\global\chapflag=1%
               \global\glgnum=0\vskip-8pt\line{\hrulefill}\mn}
\def\glg{{{\global\advance\glgnum by1}{(\number\kapnum .\number\glgnum )}}}
\def\mkglg#1{\glg\xdef#1{(\the\kapnum .\the\glgnum)}}
\def\newref#1#2{\xdef#1{#2}}
\def\ref#1{{[{\it #1\/}]}}
%%%%%%%%%%%%%%%%%%%%%%%%%%%%%%%%%%%%%%%%%%%%%%%%%%%%%%%%%%%%%%%%%%%%%%%%%%%%
\def\aifo{{\tilde a}}%{(a_i)_i^{}}
\def\cha{{\rm ch}}\def\Tr{{\rm Tr}}\def\mod{{\rm mod\ }}
\def\id{{\rm id}}

\def\cI{{\cal I}}\def\cK{{\cal K}}
\def\cO{{\cal O}}\def\cP{{\cal P}}
\def\cQ{{\cal Q}}\def\cS{{\cal S}}

\def\lb{\left(}\def\rb{\right)}
%%%%%%%%%%%%%%%%%%%%%%%%%%%%%%%%%%%%%%%%%%%%%%%%%%%%%%%%%%%%%%%%%%
\magnification=\magstep1
\font\refsl=cmcsc8
\font\refit=cmti8
\font\refrm=cmr8
\font\refbf=cmbx8
\ifnum\vers=1
\input psfig
\fi
\hoffset=-5 true mm
\font\HUGE=cmbx12 scaled \magstep4
\font\Huge=cmbx10 scaled \magstep4
\font\namen=cmr10 scaled \magstep1
\font\klein=cmr7 scaled \magstep1
\font\ttlalt=cmcsc10 scaled \magstep3 
%%%%%%%%%%%%%%%%%%%%%%%%%%%%%%%%%%%%%%%%%%%%%%%%%%%%%%%%%%%%%%
% Title Page
%%%%%%%%%%%%%%%%%%%%%%%%%%%%%%%%%%%%%%%%%%%%%%%%%%%%%%%%%%%%%%
\footline{\ifnum\pageno>0{\hss\tenrm\folio\hss}\else{\hfill}\fi}
\pageno=0
\ifnum\vers=1
\font\eurm=eurm10 scaled 2900
\font\GIANT=cmr10 scaled 2900
\setbox0=\hbox{\GIANT.\hskip2.5mm.}
\setbox1=\hbox{\eurm A}
\centerline{\eurm U\kern0.8true mm N\kern0.8true mm I\kern0.8true mm 
V\kern0.8true mm E\kern0.8true mm R\kern0.8true mm S\kern0.8true mm 
I\kern0.8true mm T\kern0.8true mm 
\raise\ht1\hbox{\lower\ht0\hbox{\GIANT.\hskip2.5mm.}}%
\hskip-\wd0A\kern0.8true mm T\kern8true mm
B\kern0.8true mm O\kern0.8true mm N\kern0.8true mm N\kern0.8true mm}
\vskip 5 true mm
\centerline{\eurm P\kern0.1true mm h\kern0.1true mm y\kern0.1true mm 
s\kern0.1true mm i\kern0.1true mm k\kern0.1true mm a\kern0.1true mm 
l\kern0.1true mm i\kern0.1true mm s\kern0.1true mm
c\kern0.1true mm h\kern0.1true mm e\kern0.1true mm s\kern3true mm
I\kern0.1true mm n\kern0.1true mm s\kern0.1true mm t\kern0.1true mm
i\kern0.1true mm t\kern0.1true mm u\kern0.1true mm t\kern0.1true mm}
\else
\vskip1cm
\centerline{\HUGE Universit\"at Bonn}
\vskip10pt
\centerline{\Huge Physikalisches Institut}
\fi
%\vskip 2.8cm
\vskip1cm
\centerline{\bf Physics Letters B 350 (1995) 203 -- 211}
\vskip1cm
%%%%%%%%%
\centerline{\ttlalt Steps towards Lattice Virasoro Algebras:}
\vskip 6pt
\centerline{\ttlalt su(1,1)}
\vskip 1.4cm
\centerline{\namen M.\ R\"osgen\raise7pt\hbox{\klein 1},
                   R.\ Varnhagen\raise7pt\hbox{\klein 2}}
\vskip 1.2cm
\centerline{\bf Abstract}
\vskip3true mm
\noindent
An explicit construction is presented for the action of the su(1,1)-subalgebra
of the Virasoro algebra on path spaces for the c(2,q) minimal models.
In the case of the Lee-Yang model, 
we show how this action already fixes the central charge of the full 
Virasoro algebra.
For this case, we additionally construct a representation in terms 
of generators of the corresponding Temperley-Lieb algebra.
\vskip 3pt
\line{\hbox to 5cm{\hrulefill} \hfill}
\line{${}^{1}$ email: roesgen@avzw02.physik.uni-bonn.de\hfill}
\line{${}^{2}$ email: raimund@avzw02.physik.uni-bonn.de\hfill}
\vskip1mm
\noindent
\ifnum\vers=1
\line{\hfill\psfig{figure=is.ps}\hfill }
\vskip -95pt
\else
\vskip1cm
\fi
\settabs \+&  \hskip 110mm & \phantom{XXXXXXXXXXX} & \cr
\+ & Post address:                         & BONN--TH--94--23& \cr
\+ & Nu{\ss}allee 12 &    \ifnum\vers=0{hep-th/9501005 }\fi & \cr
\+ & D-53115 Bonn                          & Bonn University & \cr
\+ & Germany                               & December 1994& \cr
\vfill\eject
%%%%%%%%%%%%%%%%%%%%%%%%%%%%%%%%%%%%%%%%%%%%%%%%%%%%%%%%%%%%%%%%%%%%%%
%%%%%%%%%%%%%%%%%%%%%%%%%%%%%%%%%%%%%%%%%%%%%%%%%%%%%%%%%%%%%%%%%%%%%%
\newref\ABF{ABF}\newref\Ba{Ba}\newref\Be{Be}\newref\BPZ{BPZ}
\newref\Boe{B\"o}\newref\CE{CE}\newref\JM{JM}\newref\BM{BM}
\newref\FF{FF}\newref\FGV{FGV}\newref\FNO{FNO}\newref\IT{IT}
\newref\Ka{Ka}\newref\Ke{Ke}\newref\KKMM{KM}\newref\KR{KR}
\newref\KRV{KRV}\newref\KS{KS}\newref\Me{Me}\newref\FQ{FQ}
\newref\MS{MS}\newref\N{Na}\newref\NRT{NRT}\newref\ReI{Re2}
\newref\Pa{Pa}\newref\Re{Re1}\newref\Ri{Ri}\newref\Ga{FG}
\newref\R{R\"o}\newref\VS{Sc}\newref\Jo{Jo}\newref\FV{FV}
\newref\FS{FS}\newref\AHY{AHY}\newref\BLS{BLS}\newref\WP{WP}
\def\lag{\langle}\def\rag{\rangle_{\cP}}\newref\Oc{Oc}\newref\GG{GG}
\def\nn{\vskip3pt}
\def\tip#1{{\tt #1}}\def\ti#1{}
\ifnum\vers=1
\vsize=21.5cm\voffset=-0.75cm
\fi
\ifnum\vers=2
\vsize=21cm\voffset=-0.75cm
\fi
\ifnum\vers=0
\baselineskip=16pt
\fi 
%%%%%%%%% (1) %%%%%%%%%%%%%%% Introduction %%%%%%%%%%%%%%%%%%%%%%%%%%%%
\chapter{1. Introduction}
Characters of the Virasoro minimal models \ref{\BPZ}
are now for a decade known to
correspond to spectra of CTM (corner transfer matrix) hamiltonians
for certain 2d statistical models \ref{\ABF}.
Recently, there have been some advances in the algebraic understanding 
of that relation \ref{\JM}.
However, the geometric meaning of the Virasoro algebra
in these lattice models is still unknown.
To achieve a better understanding of the latter, an explicit 
construction of the Virasoro operation on the corresponding lattice
or, as a first step, on the underlying path space seems to be promising.  
A natural starting point for such a construction are the path spaces
directly obtained from the Virasoro irreducible representation module
structure \ref{\KR,\KRV}, which in the cases we consider here 
are isomorphic to the 1d CTM-configuration spaces 
of the corresponding statistical models \ref{\Ri,\KR}. 
Some recent results on the relation of statistical models and Virasoro
character identities can be found e.g.\ in refs.\ 
\ref{\Be,\BM,\BLS,\FQ,\GG,\WP};
different approaches to lattice Kac-Moody and Virasoro algebras have
been considered by refs.\ \ref{\Ga,\FV} and \ref{\CE,\IT,\KS}.\sn
Some importance of path space structures also arises from        
the algebraic approach to conformal field theory:
It provides a natural structure for the computation of fusion rules \ref{\Re},
the explicit construction of observable and field algebras \ref{\VS},
and finally of quantum symmetries \ref{\ReI}. 
Such constructions have already been given for the Ising model \ref{\MS,\Boe} 
and some WZW-models \ref{\FGV}. \nn
Here, we consider the next simplest cases of Virasoro minimal models,
which are not contained in the unitary series (which would
be somewhat more natural from the point of view of algebraic quantum 
field theory), but the nonunitary minimal series of the Virasoro algebra
$$[L_m,L_n]=(n-m)L_{m+n}+{c\over{12}}(n^3-n)\delta_{m+n,0},\eqno\glg$$
with central charges $c(2,2K+3)$, where
$$ c(2,2K+3)=-{{2K(6K+5)}\over{2K+3}}. \eqno\glg $$
The lowest weight representations of these models have lowest weights
$$
h=h^{(2,2K+3)}_{1,m} = -{{(m-1)(2K+2-m)}\over{2(2K+3)}},
\qquad 1\leq m\leq K+1. \eqno\glg
$$
Actually, the relative simplicity of these cases in comparison to the
unitary ones lies in the simpler path space structure 
of the superselection sectors \ref{\KR,\KRV} or, in other words, 
in the simpler structure of the annihilating ideal \ref{\FNO,\FF,\FS}.\bn
More specifically,
the path representation spaces for the c(2,2K+3)-Virasoro minimal models 
are well known \ref{\JM,\FNO,\KR} to be generated by the following set
of sequences $\tilde m=(m_j)_{j\geq 0}$
taking values in $\{ 0,...,K\}$, ending at zero,
having initial value $m_0$ and obeying an additional constraint 
which is reminiscent
of the null state structure of the model \ref{\FNO}:
$$
{\cal S}(K,m_0) = \left\{ \tilde m %(m_j)_{j\geq 0}
\in \{ 0, \cdots , K\}^{\BN} \mid m_j=0 \hbox{\rm \ for }j\gg0; %m_0=m, 
m_j+m_{j+1}\leq K\ \forall j \right\}. \eqno\mkglg\pathspace
$$ 
$\cal S$ is usually described as a set of paths running on graphs
with nodes carrying labels $m_j$.
The path space $\cP(K,m)$ of formal linear combinations of elements of 
$\cS(K,K+1-m)$ after completion becomes a pseudo Hilbert space 
(not requiring positive definiteness of the --- therefore pseudo --- 
scalar product) for any choice of a non-degenerate 
pseudo scalar product in which all generating paths are orthogonal.
In the following, we will make use of that bilinear form 
$(\cdot,\cdot)\equiv \lag\cdot,\cdot\rag$
in which they become ortho{\it normal.\/}
Equipped with $L_0$ operating in a way 
motivated by the structure of corner transfer matrix hamiltonians
\ref{\Ba}, namely diagonally in the path basis \ref{FNO,KR} by
$$ L_0 \tilde m = \left(\sum_{k\geq 0} k m_k\right) \tilde m, \eqno\glg $$
it reproduces the correct Virasoro characters
$$
q^{ {c\over{24}}-h_{1,m}^{(2,2K+3)} }\chi^{(2,2K+3)}_{1 ,m}(q)=
\cha^{(2,2K+3)}_{1 ,m}(q)=q^{-h_{1,m}^{(2,2K+3)}}\Tr_{\cP(K,m)}q^{L_0}
=\Tr_{V(c,h)}q^ {L_0},
$$
which directly follows from the sum expressions of their characters
\ref{\FNO,\NRT}
$$
\eqalignno{
\cha^{(2 ,2K+3)}_{1 ,m}(q)
&= \prod_{{l\geq1,}\atop{l\not\equiv 0,\pm  m\ \mod{ (2K+3)}}} 
(1-q^l)^{-1} & \cr
&= \sum_{n_1,...,n_K \geq 0}
{{q^{N_1^2+\cdots + N_K^2 + N_{m}+\cdots + N_K }}
\over{(q)_{n_1} \cdots (q)_{n_K} }} &\mkglg\sumform}
$$
for the superselection sector with lowest weight $h_{1,m}^{(2,2K+3)}$,
where $(q)_n = (1-q)\cdots(1-q^n)$ and
$N_i = \sum_{j=i}^K n_j$. \bn
In the following, first steps towards an implementation of an 
explicit operation of the Virasoro algebra on these path spaces
are given, starting from the simplest case: The $su(1,1)$ subalgebra
in the c(2,5)-model. For convenience, we restrict our considerations
to the vacuum sector. For general sectors, the operation remains 
basically the same; they involve, however, some additional technicalities
at the beginning of the sequence \ref{\R}. \bn 
%%%%%%%%% (2) %%%%%%%%%%% Vir(2,5) %%%%%%%%%%%%%%%%%%%%%%%%%%%%%%%%%
\chapter{2. su(1,1) for the Lee Yang edge singularity c(2,5)}
\def\tabtext{
The paths $(m_j)_j\in\cP(1,m)$ generating the c(2,5)-Hilbert space
just consist of zeros and ones.
Hence, the states of lowest energy $E$ ($L_0$-eigen\-value)
in both sectors of that theory are given by the sequences in table 1 --
in agreement with the beginning of the two characters:}
\def\tabchar{\hfill\break
$\cha^{}_{1,1}(q)= 1 + q^2 + q^3 + q^4 + q^5 + 2 q^6 + ...$
and\hfill\break
$\cha^{}_{1,2}(q)= 1 + q + q^2 + q^3 + 2 q^4 + 2 q^5 + 3 q^6 + ...$\vfill
\phantom{x}\break}
\def\dispcha{
$$\cha^{}_{1,1}(q)= 1 + q^2 + q^3 + q^4 + q^5 + 2 q^6 + ...
\quad{\rm and}\quad
\cha^{}_{1,2}(q)= 1 + q + q^2 + q^3 + 2 q^4 + 2 q^5 + 3 q^6 + ...$$}
%%%End(Defs(Tabtext))
\ifnum\vers=0\tabtext\dispcha\fi
%% TABELLE
\def\tablerule{\noalign{\hrule}}
\def\smtabskip{height2pt&\omit&&\omit&&\omit&\cr}
\hbox to\hsize{
\vbox{
\begingroup\ifnum\vers>0\hsize=4.3cm\else\hsize=2cm\fi
{\hbox to \hsize{\vbox{ % Begin(text)
\ifnum\vers>0\tabtext\tabchar\fi
}}}
% End(text)
\endgroup
\vfill
} \hfill\hskip 0.15cm
\vbox{\hbox{\
\vbox{  % Begin(Tab)
\vbox{\offinterlineskip
\hrule
\halign{&\vrule#&\strut\quad\hfil#\quad\cr
\smtabskip
\smtabskip
&E\hfil&&$h_{1,1}=0$\hfil&&$h_{1,2}=-1/5$\hfil&\cr
\smtabskip
\tablerule
\smtabskip
\smtabskip
&0&&(1,0,0,0,0,0,0,0,0,0,...)&&(0,0,0,0,0,0,0,0,0,0,...)&\cr
\tablerule
&1&& -\qquad                 &&(0,1,0,0,0,0,0,0,0,0,...)&\cr
\tablerule
&2&&(1,0,1,0,0,0,0,0,0,0,...)&&(0,0,1,0,0,0,0,0,0,0,...)&\cr
\tablerule
&3&&(1,0,0,1,0,0,0,0,0,0,...)&&(0,0,0,1,0,0,0,0,0,0,...)&\cr
\tablerule
&4&&(1,0,0,0,1,0,0,0,0,0,...)&&(0,0,0,0,1,0,0,0,0,0,...)&\cr
& &&                         &&(0,1,0,1,0,0,0,0,0,0,...)&\cr
\tablerule
&5&&(1,0,0,0,0,1,0,0,0,0,...)&&(0,0,0,0,0,1,0,0,0,0,...)&\cr
& &&                         &&(0,1,0,0,1,0,0,0,0,0,...)&\cr
\tablerule
&6&&(1,0,0,0,0,0,1,0,0,0,...)&&(0,0,0,0,0,0,1,0,0,0,...)&\cr
& &&(1,0,1,0,1,0,0,0,0,0,...)&&(0,1,0,0,0,1,0,0,0,0,...)&\cr
& &&                         &&(0,0,1,0,1,0,0,0,0,0,...)&\cr
\tablerule
&...&&...\qquad&&...\qquad&\cr
\smtabskip}
\hrule}
Table 1: lowest states for $c=-22/5$
}  % Eof(Tab)
}} % of vbox(hbox)
}  % of hbox
%%% EOF(TABELLE)
\sn
Therefore, the definition of $L_0$ suggests that $L_1$ 
-- which we now want to construct --
should basically shift a single 1 in the sequence 
to the right by one step, for example
$$
\eqalign{
L_1 (1,0,0,1,0,0,0,1,0,1,0,0,0,...) 
&= \alpha_1 (1,0,0,0,1,0,0,1,0,1,0,0,0,...)\cr
&+ \alpha_3 (1,0,0,1,0,0,0,1,0,0,1,0,0,...)\cr} 
$$
(the coefficient $\alpha_2$ for the shift of the third 1 being zero
due to the obstruction given by the fourth 1 --- see eq.\ \pathspace)
and $L_{-1}$ should do the same job in the
reverse direction. This will already ensure the two commutators
$$ [L_0,L_{\pm1}] = \pm L_{\pm1}, $$ to be fulfilled
such that it only remains to adjust coefficients $\alpha$ in order to
obtain the third commutator, $[L_{-1},L_1]=2L_0,$ which will be our aim
for the remaining part of this section. \sn
In order to get a somewhat natural operation, we make use of the
following assumption:
\parindent=10pt\sn
\item{$\bullet$} $L_{\pm1}$ should act only ``locally'' on the path,
which means that it should be a linear combination of shifts
of any single 1 in the sequence, weighted by coefficients $\alpha$
which only depend on the position of the shifted 1 and global quantities
(as the number of 1's to the left or the right of the shifted 1
in the sequence, and the total number of 1's), but not on where
exactly all the other 1's are located in the sequence. 
In particular, it should not make any difference whether 
by a successive application of $L_n$-operators first 
some specific 1 and than another 1 is shifted or vice versa. Therefore,
this requirement already ensures $[L_{-1},L_1]$ to act diagonally
on $\cP$ in the path basis. Finally, the locality assumption 
ensures well-definedness of the Temperley-Lieb expressions in
section 4, and is motivated by the relation of ``path-locality'' to 
locality in the underlying model configuration space.
\sn\noindent
Now, we necessarily have to fulfill a physical requirement:\sn
\item{$\bullet$} $L_{\pm1}$ of course are not allowed to change the
superselection sector; therefore, they have to keep the beginning
of a path fixed.
\sn\noindent
For a convenient handling of special configurations in the path,
we additionally require:\sn
\item{$\bullet$} If an obstruction against the shift of a 1 in the
path occurs --- which is a pattern of type $(...,1,0,1,...)$
such that neither the left 1 is allowed to be shifted to the right
nor the right 1 to the left --- the two missing contributions
to the commutator $[L_{-1},L_1]$ should just cancel each other,
such that its total value is still fixed to the correct value of $2L_0$.
This requirement can easily be fulfilled by the ansatz
\def\lag{\langle}\def\rag{\rangle_{\cP}}
$$
\lag (...,1_i,...),L_{-1}(...,1_{i+1},...)\rag
\lag (...,1_{i+1},...),L_1(...,1_i,...)\rag = f(i-K_i,K_\infty),\eqno\glg
$$
where $(...,1_{i(+1)},...)$ is a sequence with a 1 at position $i(+1)$,
$K_i$ 1's to the left of position $i$ and containing totally $K_\infty$
1's (note that due to the locality of the shift operations, they leave
invariant all $K_i$'s evaluated on any path), 
and where $f$ is a polynomial function.
Therefore, the effect of the commutator $[L_{-1},L_1]$ on the 1 at
position $i$ gives a diagonal contribution of 
$$
\eqalign{
&\lag (...,1_i,...),L_{-1}(...,1_{i+1},...)\rag
\lag (...,1_{i+1},...),L_1(...,1_i,...)\rag\cr
&-\lag (...,1_i,...),L_1(...,1_{i-1},...)\rag
\lag (...,1_{i-1},...),L_{-1}(...,1_i,...)\rag\cr
&=f(i-K_i,K_\infty)-f(i-1-K_i,K_\infty).}\eqno\mkglg\contrib $$
\sn\parindent=0pt
Now, we have to put everything together to get the family of 
$su(1,1)$-operations which is fixed by these requirements.
In the following explicit formulas, we restrict our considerations
to the vacuum sector ($m=1$).
First of all, note that $f$ can only be of degree two as a polynomial 
in $i$, since higher degree $d$ polynomials will leave 
the commutator $[L_{-1},L_1]$  with terms
$i^{d-1}$ as contribution from the shift of the 1 at position $i$ 
up and down (and vice versa), 
which due to the locality of the operation cannot be compensated
generically. 
Thus, $d=2$, since what we like to obtain from the 
commutator is $2L_0$ which is linear in the position $i$. \sn
Summing \contrib\ over all positions $i$ with a 1, and requiring this to equal 
$2L_0 {\tilde a} = \left(2 \sum_i ia_i\right) {\tilde a}$, together with the 
initial condition $f(0,K_\infty)=0$ fixes $f$ to be
$$ f(j-K_j,K_\infty)= (j+K_\infty-K_j+2)(j-K_j), \eqno\glg $$  
which essentially consists of the position $\pm$ the number of 1's in 
the sequence to the left or the right.
We can now arbitrarily factorize $f$ into a part for $L_1$ and one for
$L_{-1}$; any choice will just fix another pseudo scalar product on the
path Hilbert space. The nicest choices are either the symmetric
one or the one just making use of the above factorization into 
polynomials of degree 1; the first one will be used for the remaining
part of this section, the other one in the next section. \sn
We therefore have derived a family of $su(1,1)$ representations on $\cP(1,1)$
in terms of counting operators $K_i, K_\infty$ and shift operators:
Using shift operators defined by
$$
S^j_+ (...,1_j,0,0,...) = (...,0_j,1,0,...) \qquad\qquad
S^j_- (...,0,0_j,1,...) = (...,0,1_j,0,...)
$$
as long as the image path is allowed by \pathspace, and operating 
identically zero otherwise, the operators
$$
L_{\pm1} := \sum_{j\geq1} L_{\pm1}^{[1];j}\qquad{\rm with}\qquad
L_{\pm1}^{[1];j} := \sqrt{(j+{K_\infty}-K_j+2)(j-K_j)} S^j_\pm\eqno\glg
$$
are just the most symmetric representative of this family of $su(1,1)$
representations, fulfilling
$$ [L_0,L_{\pm1}] = \pm L_{\pm1}\qquad\qquad [L_{-1},L_1] = 2L_0. $$
Another version will be given along with the formulas for the more 
general case $\cP(K,1)$ treated in the next section. The $h=-1/5$ sector
can essentially be treated in an analogous way; however, the commutator
$[L_{-1},L_1]=2L_0$ introduces $h$ on the right hand side of the
equations for the coefficients, such that the latter explicitly contain
$h$ \ref{\R}; furthermore, the number of 1's in an irreducible $su(1,1)$
lowest weight representation is no longer constant in generic sectors
due to an additional contribution on the beginning of the path
\ref{\R,\Ka}.\sn
Fortunately, at least for the $c(2,5)$ model $\cP$ is even ``slim enough'' 
to demonstrate that also $L_2$ is already determined, fixing the central
charge $c$ to $-22/5$.
This can be shown by inspection of the commutators 
on the states of the lowest four energy levels:
Taking $\alpha,\beta$ to be the coefficients of $L_2$, which can be 
chosen symmetrically as those of $L_1$ are, and which are to be
determined by the Virasoro commutators, we have
$$
\eqalign{
&(1,0,0,0,0,0,...)
\mathop{\longmapsto}\limits^{{1\over\alpha}L_{2}} (1,0,1,0,0,0,...)
\mathop{\longmapsto}\limits^{{1\over\alpha}L_{-2}} (1,0,0,0,0,0,...)
\mathop{\longmapsto}\limits^{L_{-2}} 0\cr
&(1,0,1,0,0,0,...)
\mathop{\longmapsto}\limits^{{1\over\beta}L_{2}} (1,0,0,0,1,0,...)
\mathop{\longmapsto}\limits^{{1\over\beta}L_{-2}} (1,0,1,0,0,0,...) \cr
&(1,0,0,0,1,0,...)
\mathop{\longmapsto}\limits^{{1\over\sqrt{10}}L_{-1}} (1,0,0,1,0,0,...) \cr
&(1,0,1,0,0,0,...)
\mathop{\longmapsto}\limits^{{1\over2}L_{1}} (1,0,0,1,0,0,...), \cr
}\eqno\glg
$$
and the commutators
$$ [L_0,L_2]=2L_2\quad [L_{-1},L_2]=3L_1\quad [L_{-2},L_2]=4L_0+c/2 $$
fix $\alpha,\beta$ and $c$ to be
$$ \beta=3\sqrt{3/5} \qquad  c=2\alpha^2=-22/5. \eqno\glg $$
Therefore, our construction of the $su(1,1)$-subalgebra
passed the natural consistency check on 
extensibility to the full Virasoro algebra.
However, an explicit construction of $L_{\pm2}$ --- and therefore, by 
computing commutators, also for the full Virasoro algebra ---
is much more involved than the one of $L_{\pm1}$.
In particular, $L_{\pm2}$ has a part which does not commute with $K_\infty$, 
and it is actually just this part of $L_{\pm2}$ which fixes the central 
charge $c$. Therefore, the situation here is in total analogy to that of
the Ising model \ref{\CE,\KS}, which however does not involve additional 
counting operators $K_i$ \ref{\R} and in that sense is less complicated. \sn
Another approach to the construction of both $L_{\pm1}$ and $L_{\pm2}$ 
for the $c(2,5)$ model was given in 
reference \ref{\Ka}. However, the consistency of that construction
has only been shown for subspaces of sufficiently low energy. 
Yet another possibility of defining a Virasoro operation on paths is
given implicitly by the basis of ref.\ \ref{\FNO}; however, the
resulting operation is not local on the paths.\bn 
%%%%%%%%% (3) %%%%%%%%%%% Vir(2,q) %%%%%%%%%%%%%%%%%%%%%%%%%%%%%%%%%
\ifnum\vers=0\vfill\eject\fi
\chapter{3. su(1,1) for the c(2,2K+3)-series}
The sum form of characters \sumform\ 
can be interpreted as the partition function for $K$ different types
of equivalence classes of local patterns in paths which for convenience will be 
called ``quasiparticles'' in the following 
(see e.g.\ \ref{\KKMM,\Me,\KRV,\Be,\BM}), and which can be excited 
independently:
The summation indices $n_i$ correspond to the number of particles of type 
(or weight) $i$ in a given configuration, 
and the $(q)_{n_i}$ in the denominator generate the combinatorics for
independent excitations in steps of 1. The $q$-exponent 
in the numerator is just the minimal energy for a configuration of 
$n_1,...,n_K$ quasiparticles of weights $1,...,K$, which corresponds to 
the ground state of the fixed particle number sector.
That ground state is for the $h_{1,m+1}$-sector ($0\leq m\leq K$) of the 
c(2,2K+3) model given by
$ (K-m)\&(m,K-m)^{n_K}\&(m,K-1-m)^{n_{K-1}}\&...\&(m,1)^{n_{m+1}}\&
(m,0)^{n_m}\&(m-1,0)^{n_{m-1}}\&...\&(1,0)^{n_1}\&z_\infty.$
In that expression and for later convenience, 
a path $\tilde m\in\cP$ is considered as the concatenation
--- defined by $(a_1,...,a_r)\&(b_1,...,b_s)=(a_1,...,a_r,b_1,...,b_s)$
--- of a finite set of patterns $p_j^k$ which are sequences of length two,
$p_j^k=(j-k,k)$ being called ``quasiparticle of weight $j$'', each of
which comes in $j$ versions
$k=0,...,j-1$, and of (finite or infinite) zero 
sequences $z_j=(0,0,0,0,...,0)$ ($j$
zeros): $\tilde m=p_{j_1}^{m_1}\&z_{r_1}\&p_{j_2}^{m_2}\&z_{r_2}\&...$
The expression $(a,b)^{n}:=(a,b,a,b,a,b,...,a,b)$ denotes the $n$-fold
concatenation of the finite sequence $(a,b)$.
If one or more of the $r_i$ are zero, it may as well happen in some
special configurations that a quasiparticle is split into two parts
$p_{-,j}^k=(j-k)$ and $p_{+,j}^k=(k)$. \sn 
As before, we require $su(1,1)$ to
leave invariant the quasiparticle structure (i.e.\ the particle
numbers $n_j$); thus, the fixed particle number sector ground state becomes an
$su(1,1)$ lowest weight vector.\sn
Therefore, it is reasonable to decompose $\cP(K,1)$ --- for explicit
formulas, we again restrict our considerations to the vacuum sector ---
into a tensor product
of path spaces $\cQ(k)$ of single quasiparticle types with energy weight $k$,
$$
\eqalign{
\cQ(k):=\{\tilde a=(a_n)^{}_{n\in\BN}\in\{0,1\}^\BN \mid \ &a_0=1,\
a_n=0\ {\rm almost\ everywhere\ },\cr
&(a_i,a_j\neq0,i\neq j\Rightarrow |i-j|\geq2k)\}.\cr}\eqno\mkglg\qdef
$$
The minimal distance $2k$ between two particles in $\cQ(k)$ corresponds
to the distance of the particles in the ground state of a sector with
fixed particle numbers according to the above interpretation of the
character formula.
On each of the tensor factors, $su(1,1)$ operates in a fashion similar
to the case $\cP(1,1)$ in the previous section,
and the operation on the tensor product
only has to be modified in order to implement position restrictions
arising from particles of higher weight in the ground state.\sn
\def\kjn{K_j^{[n]}}\def\zn{K_\infty^{[n]}}\def\Spn{S_+^{[n];j}}
\def\Smn{S_-^{[n];j}}
\def\On{\cO^{[n]}}\def\xin{\Xi^{[n]}_{\phantom{x}}}
\def\Qtenle{\id_{\cQ(1)}\otimes\cdots\otimes\id_{\cQ(n-1)}\otimes}
\def\Qtenri{\otimes\id_{\cQ(n+1)}\otimes\cdots\otimes\id_{\cQ(K)}}
In the following, our strategy is to define an $su(1,1)$-operation on 
an appropriate subspace $\cQ_K\subset\bigotimes_{1\leq k\leq K}\cQ(k)$
of the tensor product and an isomorphism 
$\cI\colon\ \cP\lb K,1\rb \mathop{\longrightarrow}\limits^{\cong}\cQ_K$
which pulls that operation as well as the quasiparticle interpretation 
back to $\cP(K,1)$. 
For that purpose, we define in analogy to the previous section 
some endomorphisms acting on sequences $\aifo\in\cQ(n)$:
$$
\matrix{
&\tilde\kjn \aifo = \lb\sum_{1\leq k\leq j+1} a_k\rb \aifo\qquad\qquad
&\tilde\zn \aifo  = \lb\sum_{k\geq 1} a_k\rb \aifo\cr
&\tilde\Spn\colon\ (...,1_j,0,...) \longmapsto (...,0_j,1,...)\qquad\qquad
&\tilde\Smn\colon\ (...,0_j,1,...) \longmapsto (...,1_j,0,...).\cr }
$$
The shift operators $\tilde\Spn,\tilde\Smn$ are zero, 
if the resulting image sequence is forbidden in $\cQ(n)$. 
For later convenience, we put these into operators
acting on the tensor product $\bigotimes_{1\leq n\leq K}\cQ(n)$,
namely for $\On=\kjn,\zn,\Spn,\Smn$
$$ \On = \Qtenle\tilde\On\Qtenri $$ and\vskip-0.8cm
$$ \xin = 2n \lb1 + \sum_{n+1\leq k\leq K} {K_\infty^{[k]}}\rb. $$
\sn
The position restrictions on ``light'' particles by the ``heavier'' particles, 
which in an $su(1,1)$ lowest weight path are located closer to the beginning 
of the path, can now be implemented into the definition of $\cQ_K$ by
means of those endomorphisms:
$$
\eqalign{
\cQ_K:=\{\tilde a=\tilde a^{(1)}\otimes...\otimes&\tilde a^{(K)}
\in\bigotimes_{1\leq k\leq K}\cQ(k)\mid a^{(n)}_j=0\ {\rm for\ }
0<j<\xi^{(n)}\cr 
&{\rm and\ }1\leq n\leq K-1,\ 
{\rm where\ } \Xi^{(n)}\tilde a=\xi^{(n)}\tilde a\}\cr
}\eqno\mkglg\qkdef
$$
Alternatively, these restrictions could also be completely implemented
into the coefficients of the $su(1,1)$-operation on the original tensor
product $\bigotimes\cQ$ as an additional offset in the position
evaluation, but for technical reasons, we prefer to use $\cQ_K$ here.\sn
The isomorphism $\cI$ is now defined step by step, starting from the
$su(1,1)$ lowest weights: 
The path $(K,0,K,0,...,0,K-1,0,K-1,0,...,2,0,1,0,1,0,0,0,...)\in\cP(K,1)$
containing $n_k$ particles of weight $k=1,...,K$ is mapped to the path
$\tilde a^{(1)}\otimes...\otimes\tilde a^{(K)}\in\cQ_K$ with
$a^{(k)}_j=1$ if either $j=0$ or $2k(n_{k+1}+...+n_K)<j\leq2k(n_k+...+n_K)$ 
with $j\in2k\BN$, all other $a^{(k)}_j$ being zero.\sn
Making use of shift operators ${\hat S}_\pm^{[n];j}$ on $\cP(K,1)$
which are defined below, $\cI$ is completely defined by the relation
$\cI{\hat S}_\pm^{[n];j}=S_\pm^{[n];j}\cI$.
By definition, the $S_\pm^{[n];j}$ of different particle index $n$
commute, and so the ${\hat S}_\pm^{[n];j}$ do. Therefore without loss
of generality, any path configuration can now be obtained
from the corresponding lowest weight by first shifting the ``light''
quasiparticles (which are most right in the ground state)
to their actual position to the right, and then to do the same for
more and more ``heavy'' ones.\sn
In order to complete the definition of $\cI$ we have to define
the operation of the ${\hat S}_\pm^{[n];\bullet}$ 
operators on $\cP(K,1)$: \sn
The shift operator ${\hat S}_+^{[j];E}$ operates on a quasiparticle
$p_j^m$ --- which has already been shifted to energy $E$ --- by
${\hat S}_+^{[j];E}(...\&z_{n_k}\&p_j^m\&z_{n_{k+1}}\&...)
=...\&z_{n_k}\&p_j^{m+1}\&z_{n_{k+1}}\&...$ 
as long as $m<j$ and $n_k,n_{k+1}>0$.
Furthermore, there are obvious rules $p_j^j\&z_n=z_1\&p_j^0\&z_{n-1}$ 
and $z_{n_1}\&z_{n_2}=z_{n_1+n_2}$.
Note that here the actual energy $E$ is the sum
of the original $L_0$ contribution $E_0$ 
of $p_j^m$ in the lowest weight path and
the number of applications of appropriate ${\hat S}_+^{[j];\bullet}$
on it for shifting it to the right, namely by
${\hat S}_+^{[j];E-1}{\hat S}_+^{[j];E-2}\cdots{\hat S}_+^{[j];E_0}$.\sn
For a definition of $\cI$ it only remains to define the 
operation of shifts ${\hat S}_\pm^{[j];\bullet}$ on successive pairs of
quasiparticles: Without loss of generality (see above), let $j_1>j_2$.
Then ${\hat S}_+^{[j_1];\bullet}$ 
(where $\bullet$ is appropriately chosen to act on $p_{j_1}$)
operates by\hfill\mkglg\concatrules
$$
\matrix{ 
&...\&p_{j_1}^{m_1}\&p_{j_2}^{m_2}\&... &\mapsto 
&...\&p_{j_1}^{m_1+1}\&p_{j_2}^{m_2}\&... &(m_1<j_1-j_2+m_2-1)\cr
&...\&p_{j_1}^{j_1-j_2+m_2-1}\&p_{j_2}^{m_2}\&... &\mapsto
&...\&p_{-,j_2}^{m_2}\&p_{j_1}^{j_2-m_2}\&p_{+,j_2}^{m_2}\&... & \cr
&...\&p_{-,j_2}^{m_2}\&p_{j_1}^{m_1}\&p_{+,j_2}^{m_2}\&... &\mapsto
&...\&p_{-,j_2}^{m_2}\&p_{j_1}^{m_1+1}\&p_{+,j_2}^{m_2}\&... 
&(j_2-m_2\leq m_1<j_1-m_2-1)  \cr
&...\&p_{-,j_2}^{m_2}\&p_{j_1}^{j_1-m_2-1}\&p_{+,j_2}^{m_2}\&... &\mapsto
&...\&p_{j_2}^{m_2}\&p_{j_1}^{m_2}\&... & \cr
&...\&p_{j_2}^{m_2}\&p_{j_1}^{m_1}\&... &\mapsto
&...\&p_{j_2}^{m_2}\&p_{j_1}^{m_1+1}\&... &(m_2\leq m_1<j_1).\cr
}
$$
Note that particles of equal type  $j_1=j_2$ are not allowed to change
their mutual order.
Finally, $\hat S_-$ just reverses the arrows. Successive application
of arbitrary shift operators can be transformed to the abovementioned
normal form by making use of the commutativity of shifts on different 
particle indices.\sn
Having collected all ingredients for the $su(1,1)$ representation under
construction including
the quasiparticle decomposition map $\cI$ for the 
vacuum sector $\cP(K,1)$ of the model c(2,2K+3),
we therefore get the following explicit $su(1,1)$ representation
on this path space:
$$
\eqalign{
L_{-1}^{[n];j}&=
\cI^{-1}\circ\lb j-(2n-1)\kjn - \xin + 2n \rb \Smn \circ\cI\cr
L_{1}^{[n];j}&=
\cI^{-1}\circ\lb j + (2n-1)\lb \zn-\kjn\rb  + \xin  \rb\Spn\circ\cI\cr
L_{-1}&=\sum_{1\leq n\leq K} \sum_{j\geq1}L_{-1}^{[n];j}\qquad{\rm and}\qquad
L_{1}=\sum_{1\leq n\leq K} \sum_{j\geq1}L_{1}^{[n];j}\cr
L_0&(l_k)_k= \lb\sum_{j\geq1} jl_j \rb (l_k)_k }\eqno\glg
$$
The details of the proof are essentially identical to the arguments
given in the previous section, and for that reason are omitted here.\sn 
Finally we have to construct a pseudo inner product
$\langle\cdot ,\cdot\rangle$ on $\cP$, such that $L_{\pm1}^*=L_{\mp1}$
and $\cP$ becomes pseudo Hilbert.
As long as we do not care about the operation of $L_{\pm2}$,
this inner product is not fixed on the ground state $v_0$ of any
sector with fixed quasiparticle numbers, that is, on the $su(1,1)$ 
lowest weight vectors; it is, however, fixed by $L_{\pm1}^*=L_{\mp1}$
as soon as the pseudo norms of just these ground states are chosen,
and can explicitly be constructed in 
terms of the inner product $(\cdot,\cdot)\equiv\lag\cdot,\cdot\rag$
on $\cP$ which is defined by orthonormality of all paths generating
the space,
namely by successive evaluation of the condition $L_{\pm1}^*=L_{\mp1}$
starting from the $su(1,1)$ lowest weight vector. \mn
It follows by construction that the sign of $\langle v,v\rangle$ is constant 
for any path $v$ taken from the same quasiparticle number sector.
Therefore, our result is in compatibility with Nahm's conjecture
on the signature characters \ref{\N,\Ke} for c(2,2K+3) defined by
$${\rm sign-}\cha^{(2 ,2K+3)}_{1 ,m}(q)\equiv
q^{-h_{1,m}^{(2,2K+3)}}\Tr_{\cP(K,m)} q^{L_0} 
{\rm sign}(\langle\cdot,\cdot\rangle),$$
namely that they are given by 
$$ {\rm sign-}\cha^{(2 ,2K+3)}_{1 ,m}(q)= \cK^{(2 ,2K+3)}_{1 ,m}(q,-1) $$
where
$$
\cK^{(2 ,2K+3)}_{1 ,m}(q,z)
= \sum_{n_1,...,n_K \geq 0}
{{q^{N_1^2+\cdots + N_K^2 + N_{m}+\cdots + N_K }}
\over{(q)_{n_1} \cdots (q)_{n_K} }}
z^{N_1 + ... + N_K}
\eqno\mkglg\sigcha
$$
in obvious similarity to eq.\ \sumform.
However, for a completion of the proof 
it remains to show that $L_2$ can be consistently defined with
this choice of $L_{\pm1}$. \bn
%%%%%%%%% (4) %%%%%%%%%%% TL(2,q) %%%%%%%%%%%%%%%%%%%%%%%%%%%%%%%%%%
\ifnum\vers=0\vfill\eject\fi
\chapter{4. Temperley-Lieb algebra for c(2,5)}
For $r$ the golden ratio ${\sqrt{5}-1}\over2$ --- satisfying $r^2=1-r$ --- 
we express the Virasoro generators for the c(2,5)-model vacuum sector in terms
of the corresponding Temperley-Lieb projections $e_i$, which obey 
\ref{\Ba,\Jo}  
$$
\eqalign{
e_i e_j &= e_j e_i \quad\quad | i - j | > 1\cr
e_i e_{i\pm1} e_i &= \tau e_i \quad\quad (\tau=r^2)\cr
e_i^2 &= e_i = e_i^*. \cr }\eqno\mkglg\TLArels
$$
On the path space $\cP(1,1)$, 
these generators are represented by \ref{\Oc,\Pa}
$$
\eqalign{  
e_{j+1} &(...,l_j,l_{j+1},l_{j+2},... ) =  \cr 
&\delta_{l_j,l_{j+2}} 
\left( 
\left[ \delta_{l_j,1} + r^{1+l_{j+1}} \delta_{l_j,0} \right] 
(...,l_j,l_{j+1},l_{j+2},... )
+ \delta_{l_j,0}\ r^{3/2} (...,l_j,1-l_{j+1},l_{j+2},... )
\right), \cr  
} \eqno\glg
$$
which obviously fulfill the Temperley-Lieb relations.
Explicitly, we have
$$ L_0 = \sum_{j\geq1} j L_0^{(j)}, \eqno\mkglg\lOsum $$
where the $L_0^{(j)}$ are defined recursively by
$$
\eqalign{
L_0^{(1)}&=0 \cr
L_0^{(2)}&=e_1 \cr
L_0^{(n)}&=\left( 1 - L_0^{(n-2)} - {1\over r} e_{n-1} \right)
\left( 1 - L_0^{(n-1)} \right)
\left( 1- {1\over r} e_{n-1} \right)
\left( 1 - L_0^{(n-1)} \right),\cr}\eqno\mkglg\lOn
$$
which is hermitean with respect to the adjoint in the Temperley-Lieb
path representation,
since in that representation the first bracket commutes with the product
of the three following ones.
More explicitly, we can make use of the fact that 
the Temperley Lieb algebra for $\tau=r^2$ contains an ideal \ref{\Jo}
generated by 
$$
\eqalign{ 
\left( -e_2e_3e_1e_2 + e_2e_3e_1 + e_1e_3e_2 - e_1e_3 \right) &+
r \left( e_3e_2e_1+e_1e_2e_3 - e_1e_2 -e_2e_1 -e_2e_3 -e_3e_2 \right)\cr
&+ (1-r) \left( e_1 + e_3 \right) + r e_2
+ (3r-2) \cr
}
$$
and its translates (by index shifts). Since that ideal is mapped to zero 
in the path representation, the first terms turn out to be
$$
\eqalign{
L_0^{(3)}&={1\over r} \left( e_1 e_2 + e_2 e_1 -e_1 -e_2 \right) +1\cr
L_0^{(4)}&={1\over r^3} e_2 e_3 e_1 e_2 + 
{1\over r} \left( e_1 e_3 - e_2 e_3 e_1 -e_1 e_3 e_2
\right).} \eqno\mkglg\lOexpl
$$
Similar expressions also hold for $L_{\pm1}$: After rewriting the
counting operators implicitly in terms of the Temperley-Lieb generators,
$$
K_\infty = 1 + \sum_{n\geq 2} L_0^{(n)}\qquad\qquad
K^{(n)} = 1 + \sum_{j=2}^n L_0^{(j)},
$$
the building blocks for $L_1$ are given by
$$
L_1^{(n)} = {1\over{\left( 1-r^{1/2} \right) \left( 1-r^{-1/2}\right)}}
\left(1-{1\over r} e_{n+1}\right) \left(1-{1\over{r^2}} e_n \right)
\left(1-e_{n+1}\right) L_0^{(n)},
$$
and the expressions for $L_{-1}^{(n)}$ can be obtained by hermitean
conjugation. Therefore, we finally get
$$
L_{\pm1}=\sum_{j\geq2}\sqrt{(j+{K_\infty}-K_j+2)(j-K_j)}L_{\pm1}^{(j)}\eqno\glg
$$
or analogously instead of the square root some coefficients of any other
appropriate choice of the function $f$ of coefficient products.
\bn
%%%%%%%%% (5) %%%%%%%%%%%% Discussion %%%%%%%%%%%%%%%%%%%%%%%%%%%%%%
\chapter{5. Discussion and Outlook}
We gave an explicit construction of the operation of some Virasoro generators
on the path spaces of c(2,2K+3)-Virasoro minimal models; although the full
algebra seems already to be fixed by the choice of the
$su(1,1)$-subalgebra, an explicit formula for $L_2$ is not yet known.
To go beyond the vacuum sector in our description, 
the $L_1$-operation at the beginning of the paths and the coefficients
have to be modified slightly: The quasi particle number $n_1$ is no longer 
conserved in this case \ref{\R,\Ka}. \sn
For the c(2,5)-model we additionally showed how to express the $su(1,1)$
subalgebra in terms of the Temperley Lieb Jones algebra on the
corresponding graph. However, in this case, explicit expressions for
the CTM from statistical mechanics are structurally simpler, and the
basis change in the algebra transforming one CTM expression to the
other still has to be worked out. 
\sn
For other models apart from the considered series of minimal models,
there still exist similar path spaces in some cases \ref{\KRV}, but
the explicit construction of the Virasoro operation 
--- and, more generally, of larger W-algebras --- seems to be 
more involved. At least in the cases of ref.\ \ref{\KRV} with
characters of type \sumform, the $su(1,1)$ operation 
on the corresponding paths should work in the same way.\bn
\chapter{Acknowledgements}
We are very grateful to W.\ Nahm for his instructive ideas and comments.
We also thank the members of his research group for many helpful
discussions on the subject, especially R.\ H\"ubel for comments on the
manuscript. In particular, we are grateful to A.\ Berkovich, J.\ Kellendonk,
A.\ Recknagel and M.\ Terhoeven.\nn 
M.\ R\"osgen is supported by Cusanuswerk, R.\ Varnhagen
by Deutsche Forschungsgemeinschaft (DFG).\bn
%%%%%%%%%%%%%%%%%%%%%%%%%%%%%%%%%%%%%%%%%%%%%%%%%%%%%%%%%%%%%MacRef
\def\book{\refsl}\def\tbf{\refbf}\def\tt{\refit}
\def\CMP#1{{Commun.\ Math.\ Phys.\ {\tbf #1}}}
\def\IMPA#1{{Int.\ J.\ Mod.\ Phys.\ {\tbf A#1}}}
\def\IMPB#1{{Int.\ J.\ Mod.\ Phys.\ {\tbf B#1}}}

\def\JSP#1{{J.\ Stat.\ Phys.\ {\tbf #1}}}

\def\LMP#1{{Lett.\ Math.\ Phys.\ {\tbf #1}}}
\def\MPL#1{{Mod.\ Phys.\ Lett.\ {\tbf #1}}}
\def\NPB#1{{Nucl.\ Phys.\ {\tbf B#1}}}
\def\PLB#1{{Phys.\ Lett.\ {\tbf #1B}}}
\def\PRL#1{{Phys.\ Rev.\ Lett.\ {\tbf #1}}}
\def\IM#1{{Invent.\ math.\ {\tbf #1}}}
%%%%%%%%% (6) %%%%%%%%%%%% References %%%%%%%%%%%%%%%%%%%%%%%%%%%%%%
\chapter{References}\parindent=30pt\refrm\font\it=cmti8\sn
\item{\ref\ABF} G.E.~Andrews, R.J.~Baxter, P.J.~Forrester,
  \ti{Eight-Vertex SOS Model and Generalized Rogers-Ramanujan-Type
  Identities,\/}
  \JSP{35} (1984) 193-266;
  P.J.\ Forrester, R.J.\ Baxter, 
  \ti{Further Exact Results of the Eight-Vertex SOS Model
  and Generalizations of the Rogers-Ramanujan Identities,\/}
  \JSP{38} (1985) 435-472
\item{\ref\AHY} P.F.\ Arndt, Th.\ Heinzel, C.M.\ Yung,
  \tip{Temperley-Lieb Words as Valence-Bond Ground States,\/}
  BONN-TH-94-26, cond-mat 9411085
\item{\ref\Ba} R.J.\ Baxter,
  {\tt Exactly Solved Models in Statistical Mechanics,\/}
  Academic Press, London, 1982 
\item{\ref\BPZ} A.A.\ Belavin, A.M.\ Polyakov, A.B.\ Zamolodchikov,
  \ti{Infinite Conformal Symmetry in Two-Dimen\-sional Quantum Field
  Theory,\/}
  \NPB{241} (1984) 333-380
\item{\ref\Be} A.\ Berkovich,
  \ti{Fermionic Counting of RSOS states and Virasoro Character Formulas
  for the Unitary Minimal Series $M(\nu,\nu+1)$,\/}
  \NPB{431} (1994) 315-348
\item{\ref\BM} A.\ Berkovich, B.M.\ McCoy, 
  \tip{Continued Fractions and Fermionic Representations for Characters
  of M(p,p') minimal models,\/}
  hep-th 9412030, BONN-HE-93-28, ITPSB-94-060
\item{\ref\Boe} J.\ B\"ockenhauer, 
  \tip{Localized Endomorphisms of the Chiral Ising Model,\/}
  DESY 94-116, hep-th 9407079
\item{\ref\BLS} P.\ Bouwknegt, A.W.W.\ Ludwig, K.\ Schoutens,
  \ti{Spinon bases, Yangian symmetry and fermionic representations of 
  Virasoro characters in conformal field theory,\/}
  \PLB{338} (1994) 448;
  \tip{Spinon basis for higher level su(2) WZW models,\/}
  hep-th 9412108;
  K.\ Schoutens, \ti{Yangian Symmetry in Conformal Field Theory,\/}
  \PLB{331} (1994) 335
\item{\ref\CE} A.~Connes, D.E.~Evans,
  \ti{Embeddings of U(1)-Current Algebras in
  Non-Commutative Algebras of Classical Statistical Mechanics,\/}
  \CMP{121} (1989) 507-525;
  D.E.\ Evans, \ti{$C^*$-Algebraic Methods in Statistical Mechanics
  and Field Theory,\/} \IMPB{4} (1990) 1069-1118
\item{\ref\FV} L.\ Faddeev, A.Yu.\ Volkov, 
  \ti{Abelian current algebra and the virasoro algebra on the lattice,\/}
  \PLB{315} (1993) 311-318
\item{\ref\Ga} F.\ Falceto, K.\ Gawedzki,
  \ti{Lattice Wess-Zumino-Witten Model and Quantum Groups,\/}
  J.\ Geom.\ Phys.\ 11 (1993) 251-279
\item{\ref\FF} B.\ Feigin, E.\ Frenkel, 
  \ti{Coinvariants of nilpotent subalgebras of the 
  Virasoro algebra and partition identities,\/}
  Adv.\ Sov.\ Math.\ 16 (1993) 139-148, hepth 9301039
\item{\ref\FNO} {B.L.~Feigin, T.~Nakanishi, H.~Ooguri,
  \ti{The annihilating ideals of minimal models,\/}
  \IMPA{7} Suppl.\ {\tbf 1A} (1992) 217-238}
\item{\ref\FQ} O.\ Foda, Y.H.\ Quano,
  \tip{Polynomial identities of the Rogers-Ramanujan type,\/}
  hep-th 9407191;
  \tip{Virasoro character identities from the Andrews-Bailey construction,\/}
  hep-th 9408086
\item{\ref\FS}   E.\ Frenkel, A.\ Szenes,
  \ti{Dilogarithm identities, q-difference equations and the Virasoro
  algebra,\/}
  Duke Math.\ J.\ 69 (1993) 53-60;
  \tip{Crystal bases, dilogarithm identities and torsion in algebraic
  K-groups,\/}
  hep-th 9304118
\item{\ref\FGV} J.\ Fuchs, A.\ Ganchev, P.\ Vecserny\'es,
  \ti{Level 1 WZW Superselection Sectors,\/} \CMP{146} (1992) 553-583;
  \ti{Simple WZW Superselection Sectors,\/} \LMP{28} (1993) 31-41
\item{\ref\GG} G.\ Georgiev, 
  \tip{Combinatorial construction of mudules for infinite-dimensional
   Lie-Algebras, I.\ Principal subspace,\/} hep-th 9412054
\item{\ref\IT}
  H.~Itoyama, H.B.~Thacker,
  \ti{Lattice Virasoro Algebra and Corner Transfer Matrices in the
  Baxter Eight-Vertex Model,\/}
  \PRL{58} (1987) 1395-1398;
  H.B.~Thacker, H.~Itoyama,
  \ti{Integrability, Conformal Symmetry, and Noncritical
  Virasoro Algebras,\/}
  \NPB{5A} (Proc.\ Suppl.) (1988) 9-14
\item{\ref\JM}
  M.\ Jimbo, T.\ Miwa,
  \tip{Algebraic Analysis of solvable Lattice Models,\/}
  RIMS-981, May 1994, and references therein;
  E.\ Date, M.\ Jimbo, A.\ Kuniba, T.\ Miwa, M.\ Okado,
  \ti{Exactly Solvable SOS Models: Local Height Probabilities
  and Theta Function Identities,\/}
  \NPB{290} (1987) 231;
  B.\ Davies, O.\ Foda, M.\ Jimbo, T.\ Miwa, A.\ Nakayashiki,
  \ti{Diagonalization of the XXZ Hamiltonian by Vertex Operators,\/}
  \CMP{151} (1993) 89-154
\item{\ref\Jo} V.F.R.~Jones, \ti{Index for Subfactors,\/} \IM{72} (1983) 1-25
\item{\ref\Ka} R.M.\ Kaufmann, 
  \tip{Path Space Decompositions for the Virasoro Algebra and its
  Verma Modules,\/}
  BONN-TH-94-05, hep-th 9405041
\item{\ref\KKMM} 
  R.\ Kedem, T.R.\ Klassen, B.M.\ McCoy, E.\ Melzer,
  \ti{Fermionic Sum Representations for Conformal Field Theory
  Characters,\/} \PLB{307} (1993) 68-76;
  \ti{Fermionic Quasiparticle Representations for Characters 
  of $G_1^{(1)} \times G_1^{(1)} / G_2^{(1)}$,\/} \PLB{304} (1993) 263-270
\item{\ref\KR} J.~Kellendonk, A.~Recknagel,
  \ti{Virasoro Representations on Fusion Graphs,\/}
  \PLB{298} (1993) 329-334
\item{\ref\KRV} J.~Kellendonk, M.~R\"osgen, R.~Varnhagen,
  \ti{Path Spaces and W-Fusion in Minimal Models,\/}
  \IMPA{9} (1994) 1009-1023
\item{\ref\Ke}
  A.\ Kent, \ti{Signature Characters for the Virasoro algebra,\/}
  \PLB{269} (1991) 315-318
\item{\ref\KS} W.M.\ Koo, H.\ Saleur, 
  \ti{Representations of the Virasoro algebra from lattice models,\/}
  \NPB{426} (1994) 459-504; 
  H.~Saleur, {\tt Virasoro and Temperley Lieb Algebras,\/}
  pp.~485-496 in {\book Knots, Topology and QFT,\/}
  Proceedings of the Johns Hopkins Workshop, Florence 1989,
  ed.~L.~Lusanna
\item{\ref\MS} G.\ Mack, V.\ Schomerus, 
  \ti{Conformal Field Algebras
  with Quantum Symmetry from the Theory of Superselection Sectors,\/}
  \CMP{134} (1990) 139-196;
  \ti{Quasiquantum Group Symmetry and Local Braid Relations in the 
  Conformal Ising Model,\/} \PLB{267} (1991) 207-213 
\item{\ref\Me} E.\ Melzer,
  \ti{Fermionic Character Sums and the Corner Transfer Matrix,\/}
  \IMPA{9} (1994) 1115-1136;
  \ti{The many Faces of a Character,\/}
  \LMP{31} (1994) 233-246;
  \tip{Supersymmetric Analogs of the Gordon-Andrews Identities, and
  Related TBA Systems,\/} TAUP 2211-94, hep-th 9412154
\item{\ref\N} W.\ Nahm, private communication
\item{\ref\NRT} W.\ Nahm, A.\ Recknagel, M.\ Terhoeven,
  \ti{Dilogarithm identities in Conformal Field Theory,\/}
  \MPL{A8} (1993) 1835-1848;
  M.\ Terhoeven, 
  \tip{Lift of Dilogarithm to Partition Identities,\/}
  BONN-HE-92-36, hep-th 9211120
\item{\ref\Oc} A.\ Ocneanu, \tip{Quantized groups, string algebras and
  Galois theory for algebras} in
  {\book Operator Algebras and Applications II}, 
  Lecture Notes of the London Math.\ Soc.\ vol.\ 135, 
  Cambridge University Press, Cambridge 1988
\item{\ref\Pa} V.\ Pasquier, 
  \ti{Two-Dimensional Critical Systems Labelled by Dynkin Diagrams,\/}
  \NPB{285} (1987) 162-172
\item{\ref\Re} A.~Recknagel,
  \ti{Fusion Rules from Algebraic K-Theory,\/}
  \IMPA{8} (1993) 1345-1357; 
  \tip{Ein K-theoretischer Zugang zu den Fusionsregeln konformer
  Quantenfeldtheorien,\/} thesis, BONN-IR-93-61
\item{\ref\ReI} A.\ Recknagel,
  {\tt AF-algebras and applications of K-theory in conformal 
  field theory,\/}        
  Talk given at the International Congress in Mathematical Physics,
  Satellite colloquium on ``New Problems in the General Theory
  of Fields and Particles'', 
  Paris, July 1994
\item{\ref\Ri} H.\ Riggs, 
  \ti{Solvable Lattice Models with Minimal and Nonunitary Critical
  Behaviour in Two Dimensions,\/}
  \NPB{326} (1989) 673-688
\item{\ref\R} M.\ R\"osgen, 
  \tip{Pfaddarstellungen\ minimaler\ Modelle,\/}
  Diplomarbeit, BONN-IR-93-24
\item{\ref\VS} V.\ Schomerus, 
  \tip{Construction of Field Algebras with Quantum Symmetry from 
  Local Observables,\/}
  HUTMP-93-B333, hep-th 9401042
\item{\ref\WP} S.O.\ Warnaar, P.A.\ Pearce, 
  \tip{ADE Polynomial and Rogers-Ramanujan Identities,\/}
  hep-th 9411009
\bye